\documentclass[
pra,reprint,superscriptaddress,nofootinbib]{revtex4-2}

\usepackage{times}
\def\supplementfilename{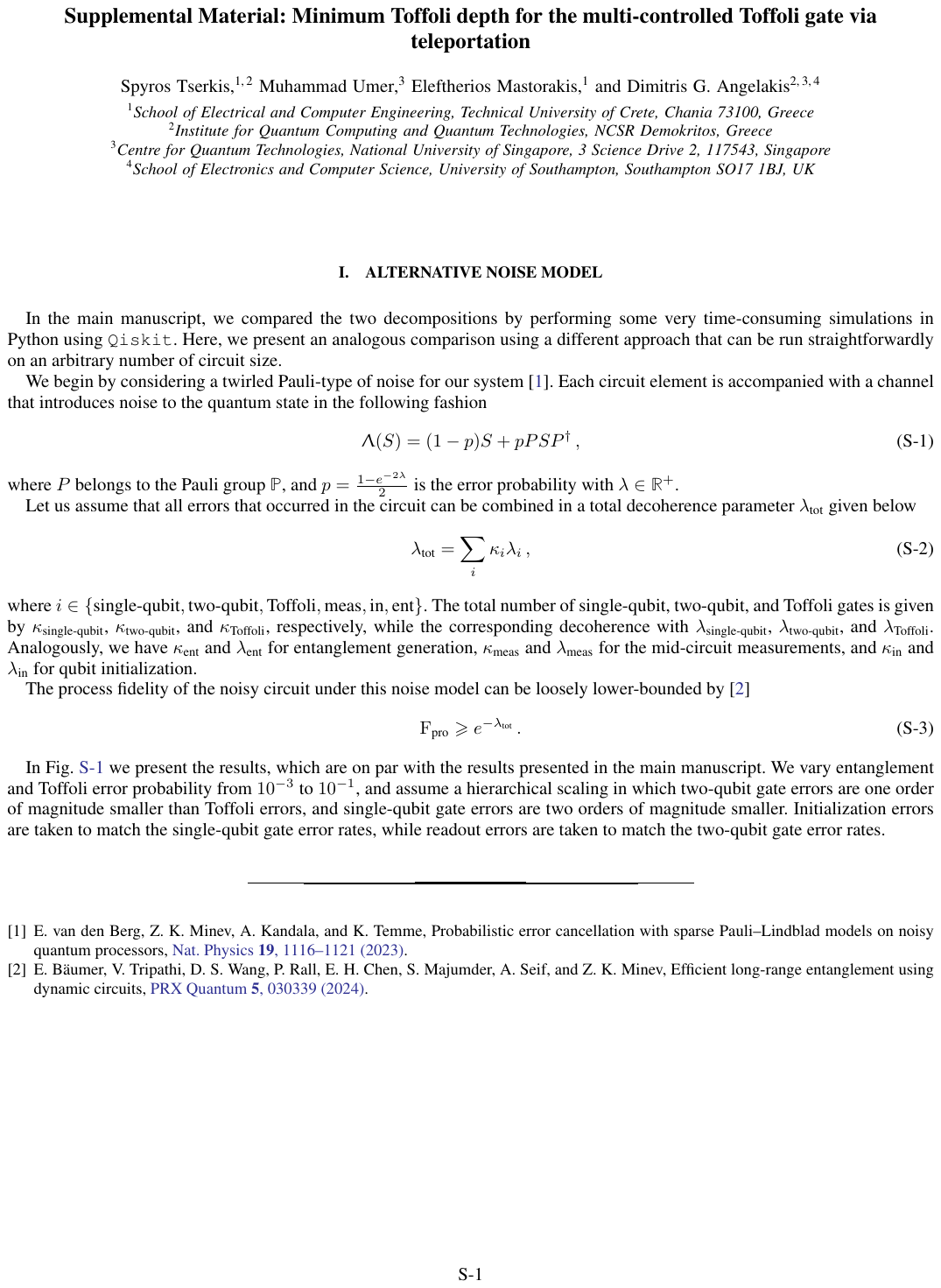}

\usepackage{fancyhdr}
\usepackage{graphicx}
\usepackage{float}
\usepackage{tabularx,booktabs} 
\usepackage{enumitem}
\usepackage{multirow}
\usepackage{adjustbox, tikz}
\usetikzlibrary{quantikz2}

\usepackage{physics}
\usepackage{amsmath}
\usepackage{amssymb}
\usepackage{dsfont}
\usepackage{mathrsfs} 
\usepackage{euscript}
\usepackage{upgreek}
\usepackage{mathtools}
\usepackage{mathdots}
\usepackage[bbgreekl]{mathbbol}
\DeclareSymbolFontAlphabet{\mathbbol}{bbold}
\DeclareSymbolFontAlphabet{\mathbb}{AMSb}

\usepackage{amsmath}
\usepackage{amsthm}
\theoremstyle{remark}

\makeatletter
\def\maketag@@@#1{\hbox{\m@th\normalfont\normalsize#1}}  
\makeatother

\usepackage[usenames,dvipsnames,table]{xcolor}
\definecolor{blue_ref}{RGB}{46,48,146}
\usepackage{colortbl}

\usepackage[colorlinks=true,urlcolor=blue_ref,citecolor=blue_ref,linkcolor=blue_ref]{hyperref}

\usepackage{comment} 
\usepackage{soul} 
\setstcolor{red}
\usepackage{cancel}
\usepackage{lipsum}

\usepackage{pgffor}

\usepackage{pdfpages} 
\makeatletter
\AtBeginDocument{\let\LS@rot\@undefined}
\makeatother

\usepackage[us,hhmmss]{datetime} 

\newcommand{\m}[1]{\mathit{#1}}
\newcommand{\f}[1]{\mathrm{#1}}
\newcommand{\map}[1]{\ifcat\noexpand#1\relax#1\else{\mathcal{#1}}\fi}
\newcommand{\set}[1]{\mathbbol{#1}}

\newcommand{\onorm}[1]{{\left\vert\kern-0.25ex\left\vert\kern-0.25ex\left\vert #1 
    \right\vert\kern-0.25ex\right\vert\kern-0.25ex\right\vert}}
    
\newcommand{\eq}[2]{\begin{equation} \label{eq:#1} #2 \end{equation}}
\newcommand{\alsub}[2]{\begin{subequations}\label{eq:#1}\begin{align} #2 \end{align}\end{subequations}}
\let\leq\leqslant
\let\geq\geqslant

\newcommand{\qs}{S}
\newcommand{\prob}{\mathbb{P}}

\usepackage[ruled,vlined]{algorithm2e}

\newcommand{\ketf}[1]{\mbox{$| #1 \rangle$}}
\newcommand{\braketf}[2]{\mbox{$\langle #1 | #2 \rangle$}}

\def\TUC{School of Electrical and Computer Engineering, Technical University of Crete, Chania 73100, Greece}

\def\NUS{Centre for Quantum Technologies, National University of Singapore, 3 Science Drive 2, 117543, Singapore}

\def\NCSR{Institute for Quantum Computing and Quantum Technologies, NCSR Demokritos, Greece}

\def\Southampton{School of Electronics and Computer Science, University of Southampton, Southampton SO17 1BJ, UK}

\begin{document}

\title{Minimum Toffoli depth for the multi-controlled Toffoli gate via teleportation}

\author{Spyros Tserkis}
\email{spyrostserkis@gmail.com}
\affiliation{\TUC}
\affiliation{\NCSR}

\author{Muhammad Umer}
\email{umer@u.nus.edu}
\affiliation{\NUS}

\author{Eleftherios Mastorakis}
\email{emastorakis@tuc.gr}
\affiliation{\TUC}

\author{Dimitris G. Angelakis}
\email{dimitris.angelakis@gmail.com}
\affiliation{\NCSR}
\affiliation{\NUS}
\affiliation{\Southampton}


\begin{abstract}
The decomposition of complex quantum operations into experimentally feasible gate sets has been a central challenge since the early development of quantum computing. The multi-controlled Toffoli (MCT) gate is a key example, with applications across a wide range of quantum algorithms, whose decomposition into smaller gates, however, typically leads to deep circuits. In this work, we introduce a teleportation-based decomposition that implements an MCT gate. This decomposition has unit Toffoli depth, independent of the number of controls, while achieving the lowest Toffoli count among existing approaches for more than five controls at the cost of a linear overhead in ancilla qubits. We also investigate the parameter regimes in which the teleportation-based decomposition is preferable by analyzing the relative quality of distributed entanglement and Toffoli gates. We further demonstrate the advantages of this implementation in circuits that rely on MCT gates, such as the adder operator, quantum read-only memory, quantum neurons, and quantum decision trees.
\end{abstract}

\maketitle

\section{Introduction}
Quantum algorithms often rely on complex operations that in practice need to be decomposed into simpler gate sets to be implemented. An example of such a complex operation is the multi-controlled Toffoli (MCT) gate, which is the generalization of the Toffoli gate over more than three qubits, or analogously, the generalization of the controlled-NOT (CNOT) gate over more than two qubits. 

MCT gates are used in a variety of quantum applications, including quantum arithmetics~\cite{Vedral_Barenco_Ekert_PRA_96, Douglas_Wang_PRA_09}, variational quantum algorithms~\cite{Lubasch2020, Jaksch2023, Umer2025, Tomesh_etal_Q_24, Over_etal_CF_25}, quantum memories~\cite{Phalak2022}, and quantum machine-learning~\cite{Tacchino2019, Cuellar2024}. Their direct physical implementation is highly challenging, so determining how to efficiently decompose it to reduce the circuit depth and other metrics, e.g., Toffoli gate count, remains an active area of research~\cite{Balauca_Arusoaie_B_22, Nemkov_etal_Q_23, Nakanishi_etal_PRA_24, Nie_Zi_Sun_arxiv_24, Kole_etal_IEEE_24, Heussen_etal_PRXQ_24, Khattar_Gidney_Q_25, Dutta_etal_PRA_25, Zindorf_Bose_PRA_25, Zindorf_Bose_Q_25}. Reducing circuit depth, often at the expense of circuit width, can be found effective in reducing the impact of decoherence during computation, which has led to increased research interest in this direction~\cite{Moore_Nilsson_SIAM_01, Anders_etal_PRA_10, Proctor_Andersson_Kendon_PRA_13, Watts_etal_ACM_19, Liu_Gheorghiu_Q_22, Buhrman_etal_Q_24, Quek_Kaur_Wilde_Q_24, Baumer_etal_PRXQ_24, Baumer_Woerner_PRR_25, Cao_Eisert_PRL_26, Tserkis_Umer_Angelakis_arxiv_25}.

In this work, we introduce a teleportation-based decomposition of the MCT gate that surpasses a recently established bound on Toffoli depth~\cite{Dutta_etal_PRA_25}. In particular, the resulting implementation achieves unit Toffoli depth, irrespective of the number of controls, and the lowest Toffoli count compared to state-of-the-art decompositions for more than five controls, at the expense of a linear increase in the number of ancilla qubits.

The proposed decomposition requires certain qubits to be entangled at the outset. Within the traditional monolithic quantum computing (MQC) paradigm, consisting of a single quantum processing unit (QPU), preparing and distributing the required entangled states is possible but impractical. By contrast, the distributed quantum computing (DQC) paradigm~\cite{Grover_arXiv_97, Cirac_etal_PRL_99, Monroe_etal_PRA_14, Caleffi2024, Barral2025, Knrzer2026}, in which multiple QPUs are linked through entanglement, can make the decomposition considerably more efficient. In DQC, entanglement is established through photon-mediated processes that have already been demonstrated with high fidelity across several quantum computing platforms~\cite{Pompili_etal_S_21, Storz_etal_N_23, Li_Thompson_PRXQ_24, Knaut_etal_N_24, OReilly_etal_PRL_24, PsiQuantum_N_25, Saha_etal_NC_25, Song_etal_NN_25, Stas_etal_N_26}.

We further consider an $\textrm{MCT}_8$ gate and investigate its optimal implementation under a given noise model. We analyze both the MQC and DQC paradigms, employing in each case the decomposition best suited to the corresponding architecture: the decomposition achieving the lower bound derived by Dutta \textit{et al.}~\cite{Dutta_etal_PRA_25} for MQC, and our teleportation-based decomposition for DQC. This comparison allows us to determine which computational paradigm provides the higher-fidelity implementation as a function of the relative noise affecting local Toffoli gates and distributed entanglement links. We also highlight the relevance of MCT decompositions for constructing adder operators, which serve as building blocks for differential operators, as well as for applications in quantum memory and quantum machine learning.

The remainder of the paper is structured as follows. In section~\ref{sec_MCT_operation} we define the MCT gate. We also present the state-of-the-art decompositions and provide a comparative analysis among them. The teleportation-based MCT decomposition is presented in section~\ref{sec_tele_decomposition}, and in section~\ref{sec_comparison} we present a comparative error analysis. In section~\ref{Sec:Applications} we discuss applications that require the use of the MCT gate. Finally, section~\ref{sec_conclusion} concludes the manuscript.

\section{Multi-Controlled Toffoli Operation}
\label{sec_MCT_operation}

\subsection{Multi-Controlled Toffoli Gate}
\label{sec_MCT_gate}

\begin{figure}[t!]
\centering
\includegraphics[width=0.7\columnwidth]{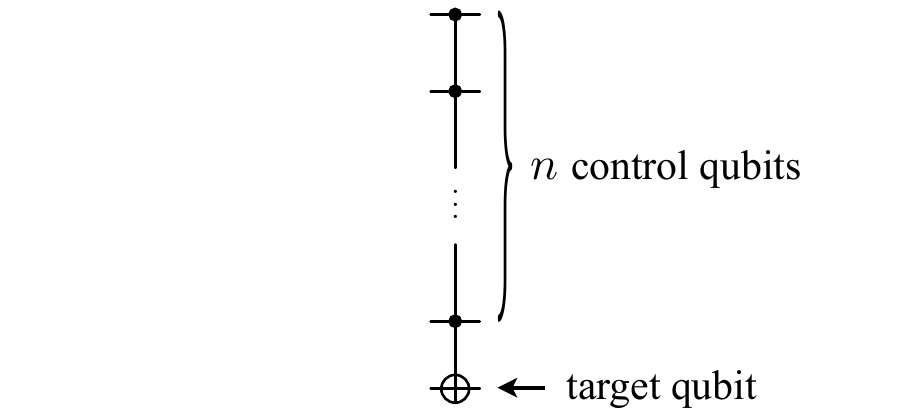}
\caption{Quantum circuit representation of an $(n{+}1)$-qubit MCT gate.}
\label{fig:MCT}
\end{figure}

\begin{figure*}[t!]
\centering
\includegraphics[width=\textwidth]{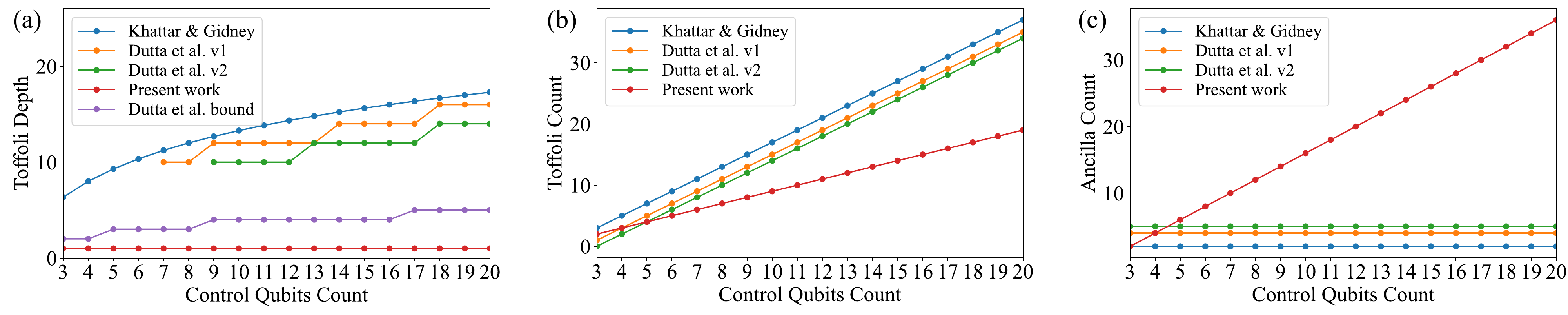}
\caption{In panels (a)-(c) we plot the Toffoli depth, the Toffoli count, and the ancilla count, respectively, against the number of control qubits in an MCT gate decomposition. The color code is the same for all three panels. With blue we have the corresponding values for the decomposition by Khattar and Gidney~\cite{Khattar_Gidney_Q_25}. With orange the corresponding values for the protocol derived by Dutta \textit{et al.}~\cite{Dutta_etal_PRA_25} that uses four ancilla qubit and with green the corresponding protocol by the same authors that uses five ancilla qubits. With red we have the corresponding values for the teleportation-based protocol proposed in this work. With purple there is a trend appearing only on panel (a) corresponding to the lower bound of Toffoli depth derived by Dutta \textit{et al.}~\cite{Dutta_etal_PRA_25}.}
\label{fig:toffoli_depth}
\end{figure*}

The $(n{+}1)$-qubit MCT gate, denoted as $\textrm{MCT}_{n+1}$, is defined by the following operation
\eq{}{
|c_1, \cdots, c_n,\, t\rangle 
\;\longmapsto\; 
|c_1,  \cdots,  c_n,\, t \oplus \bigwedge_{i=1}^{n} c_i\rangle\,,
} 
where $c_i$ corresponds to the $i$-th control qubit and $t$ to the target one. The symbol $\oplus$ in this equation indicates the XOR operation and $\wedge$ the AND operation.

In this context, the CNOT gate is equivalent to an $\textrm{MCT}_2$ gate and the Toffoli gate to an $\textrm{MCT}_3$ gate. In Fig.~\ref{fig:MCT} we present an $\textrm{MCT}_{n+1}$ gate as a quantum circuit, where the NOT operation is represented by a ``target'' symbol $\oplus$ corresponding to the Pauli $\m{X}$ matrix. Analogously, we could define gates where the target qubit is transformed according to the Pauli matrix $\m{Y}$ or $\m{Z}$. Since $\m{Y} = \m{S}^\dag \m{X} \m{S}$ and $\m{Z} = \m{H} \m{X} \m{H}$, where $\m{S}$ and $\m{H}$ (Hadamard) are typical computing gates, the conversion from one to another is trivial, so for simplicity we focus on the target matrix $\m{X}$.

\subsection{MCT Gate Decompositions}
\label{sec_decompositions}

MCT gates consisting of more than two qubits are too complex to be directly implemented in physical platforms. For that reason, they need to be decomposed into smaller gates. Various decompositions have been proposed depending on the relevant optimization. For example, one optimization is the minimization of the total number of Toffoli gates needed, referred to as Toffoli count. Another is the minimization of the circuit depth with regard to a specific gate, for instance, the Toffoli depth. Since Toffoli gates are themselves too complex for direct implementation, their decomposition can be further taken into account, e.g., the minimization of the T or CNOT count, or the corresponding circuit depth. Furthermore, gates acting on distant (i.e., non-neighboring) qubits is also a challenging task in practice, so their corresponding implementation through solely closest-neighboring qubits can also be taken into account in the analysis.

Several MCT decompositions have been suggested in the literature. In the work by Nie \textit{et al.}~\cite{Nie_Zi_Sun_arxiv_24} the MCT gate is decomposed on a circuit with a Toffoli depth equal to $20\log_2 n$ requiring a single ancilla qubit. Another MCT decomposition comes from the work by Khattar and Gidney~\cite{Khattar_Gidney_Q_25}, where the Toffoli depth is equal to $2n-3$ for a single ancilla qubit and approximately $4\log_2n$ with two ancilla qubits. In the more recent work by Dutta \textit{et al.}~\cite{Dutta_etal_PRA_25}, more decompositions were derived, alongside a lower bound of $\lceil \log_2(n) \rceil$ for the Toffoli depth, which, as we show later in this work, can be surpassed. That bound was derived under a specific framework in which the MCT operation is constructed by a computation, which encodes the conjunction of the control qubits into ancilla qubits and uses them to flip the target, followed by an uncomputation, which resets the ancilla qubits and restores the control qubits to their original state. The uncomputation part can be carried out either unitarily, which roughly doubles the Toffoli depth, or non-unitarily with the help of mid-circuit measurements, which do not increase the Toffoli depth; the Dutta \textit{et al.} decomposition use the latter approach. In contrast, our decomposition does not follow that compute-uncompute framing but a completely different one, i.e., gate teleportation, and that is why unit Toffoli depth can be reached. In Fig.~\ref{fig:toffoli_depth}(a) we present how the Toffoli depth scales over a range of twenty control qubits for different decompositions, and in Fig.~\ref{fig:toffoli_depth}(b) the corresponding Toffoli count. In Fig.~\ref{fig:toffoli_depth}(c) we compare the additional qubits, also known as ancilla qubits, required for each implementation. 

In Section~\ref{sec_comparison}, we show that Toffoli depth and count alone are insufficient to fully characterize the performance of the decompositions. A complete comparison must also account for the assumptions concerning the overall circuit depth and the noise associated with the different operations. Since the ultimate objective is the implementation of the MCT gate, we introduce a common noise model and compare the decompositions in terms of their process fidelity. As a representative case, we consider the MCT-8 gate, for which Dutta \textit{et al.} provide an explicit circuit achieving their reported bound, and compare it with the teleportation-based decomposition proposed here. For each decomposition, we adopt the computational paradigm to which it is best suited: the MQC for the decomposition of Dutta \textit{et al.} and the DQC paradigm for the teleportation-based decomposition.

\section{Teleportation-Based MCT Decomposition}
\label{sec_tele_decomposition}

\subsection{The Protocol}

\begin{figure}[b!]
\centering
\includegraphics[width=\columnwidth]{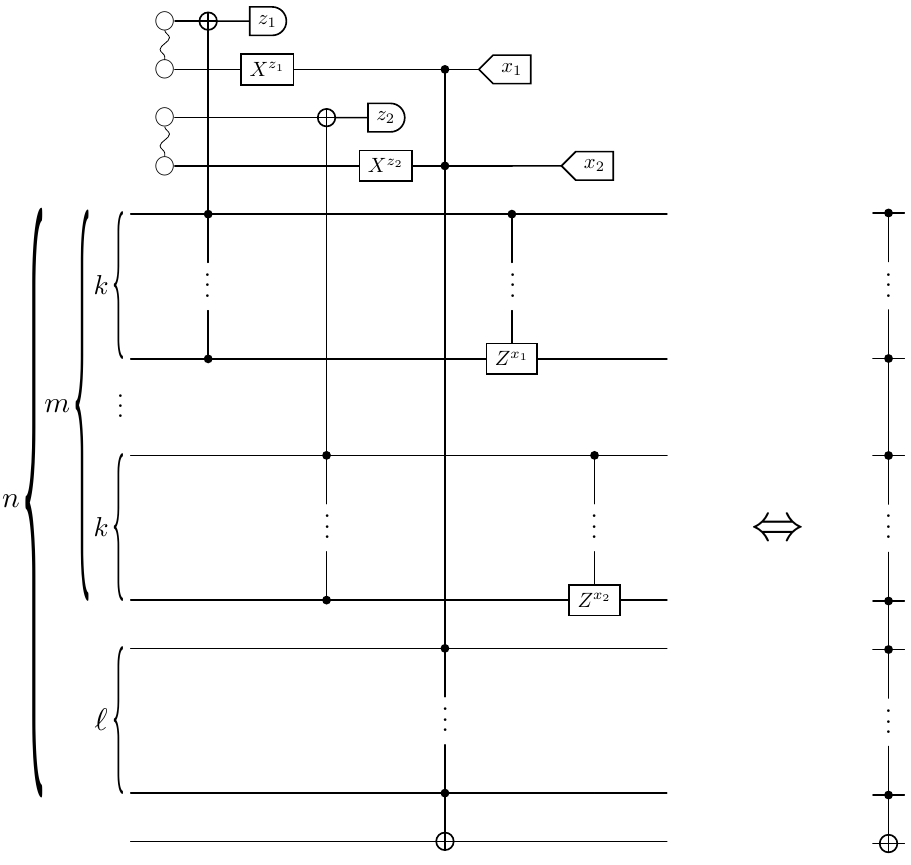}
\caption{Protocol for the teleportation of an $\textrm{MCT}_{n+1}$ gate based on Ref.~\cite{Sarvaghad_Zomorodi_QIP_21}. The $\textrm{MCT}_{n+1}$ gate is teleported using $m$ number of $\textrm{MCT}_{k+1}$ gates and a single $\textrm{MCT}_{m + \ell + 1}$ gate. The squiggly lines correspond to Bell states of the form $\ket{\Phi_+} = (\ket{00} + \ket{11})/\sqrt{2}$. Note that there are two types of measurements, in the Z and in the X basis, denoted as $z_i$ and $x_i$, respectively.}
\label{fig:teleportation}
\end{figure}

Gate teleportation~\cite{Nielsen_Chuang_PRL_97, Gottesman_Chuang_N_99, Eisert_etal_PRA_00, Sarvaghad_Zomorodi_QIP_21} is a quantum protocol capable of using entanglement as a resource to perform a gate in qubits that can be in principle spatially separated. There are multiple ways to teleport a quantum gate. For the purposes of this work, we use the protocol introduced in Ref.~\cite{Sarvaghad_Zomorodi_QIP_21} and express it in the following parametrized form: the $\textrm{MCT}_{n+1}$ gate can be teleported using a circuit with $m$ $\textrm{MCT}_{k+1}$ gates and a single $\textrm{MCT}_{m+\ell+1}$ gate, where the following equation is satisfied
\eq{condition}{
n = m k + \ell\,. 
}
The protocol also uses $2m$ measurements, half in the Z basis and half in the X basis. In total, it uses $2m$ ancilla qubits and $m$ entangled pairs for the construction of an $\textrm{MCT}_{n+1}$ gate~\footnote{It should be noted that the first paper discussing the $\textrm{MCT}_{n+1}$ gate teleportation was Ref.~\cite{Eisert_etal_PRA_00}, which is a special case of the protocol in Ref.~\cite{Sarvaghad_Zomorodi_QIP_21} for $k=1$.}. In Fig.~\ref{fig:teleportation} we present this general protocol as a circuit. 

Our goal is to teleport an arbitrary $\textrm{MCT}_{n+1}$ gate with $n \geq 2$ using only Toffoli gates. We start by setting $k=2$, so the set of $m$ $\textrm{MCT}_{k+1}$ gates corresponds to Toffoli gates. Thus, the condition in Eq.~\eqref{eq:condition} becomes
\eq{condition2}{
n = 2 m  + \ell \,,
}
where we set
\eq{}{
\ell = \left\{
\begin{array}{cl}
      0 & \text{for even} \;\; n \,, \\
      1 & \text{for odd} \;\; n \,. \\
\end{array}
\right. 
}
That implies
\eq{l_expression}{
\ell = \frac{1 - (-1)^n}{2} \,,
}
so $m$ can be written as 
\eq{m_expression}{
m = \frac{2n -1 + (-1)^n}{4} \,.
}

Then, we recursively employ the gate teleportation protocol in the following way. For the first iteration, we set $n_1 = n$, $m_1 = m$, and $\ell_1 = \ell$. In each iteration of the protocol, we replace $n_i$ by $n_{i+1}$, which is equal to
\eq{recursion}{
n_{i+1} = m_i + \ell_i \,.
}
The iteration ends at the first index for which
\eq{term_condition}{
m_i + \ell_i = 2 \,,
}
denoted as $i = i_{\text{max}}$. The only exception to this rule is the case where $n_1=2$, for which $m_1=1$ and $\ell_1 = 0$. In that case, the teleportation protocol is invoked once, so we trivially have $i_{\text{max}} = 1$. In Appendix \ref{appA} we analytically derive the expression
\eq{imax}{
i_{\text{max}} = \left\{
\begin{array}{cl}
      1 & \text{for} \;\; n = 2  \,, \\
      \left\lceil \log_2 n \right\rceil - 1 & \text{for} \;\; n \geq 3 \,. \\
\end{array}
\right. 
}

Each time the teleportation protocol is employed new ancilla qubits appear that need to be entangled with qubits from the rest of the circuit. Since entanglement does not need to be shared only among closest-neighboring qubits, we can reorder qubits in a way that the Toffoli gates appear strictly on closest-neighboring qubits. Below, we provide two concrete examples of this decomposition protocol for the $\textrm{MCT}_5$ and the $\textrm{MCT}_8$ gates.

\subsubsection{Example: Decomposing the $\textrm{MCT}_5$ gate}

\begin{figure}[t!]
\centering
\includegraphics[width=\columnwidth]{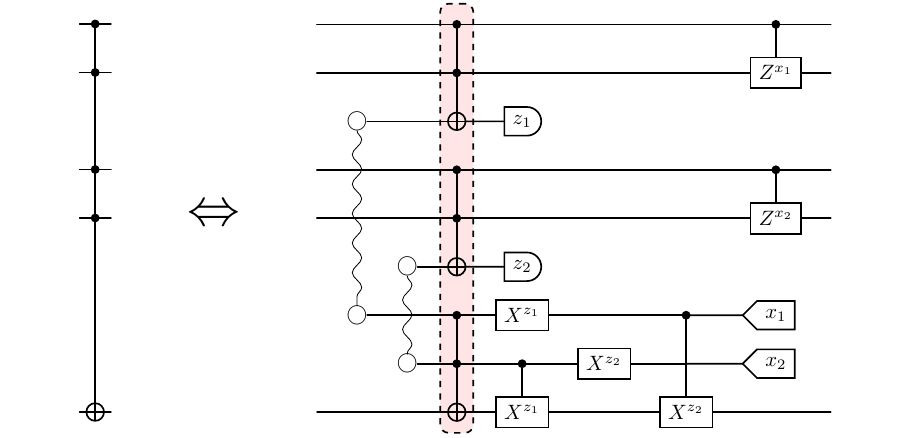}
\caption{An $\textrm{MCT}_5$ gate as a unitary operation (left) and through a teleportation-based decomposition (right).}
\label{fig:mct5}
\end{figure}

For the $\textrm{MCT}_5$ gate, we obtain the following sequence:
\alsub{}{
i = 1: \quad n_1 = 4 \quad \longmapsto  \quad m_1 = 2 \,, \quad \ell_1 = 0 \,. \nonumber
}
The corresponding circuit for this decomposition is depicted in Fig.~\ref{fig:mct5}.

\subsubsection{Example: Decomposing the $\textrm{MCT}_8$ gate}

\begin{figure*}[t!]
\centering
\includegraphics[width=\textwidth]{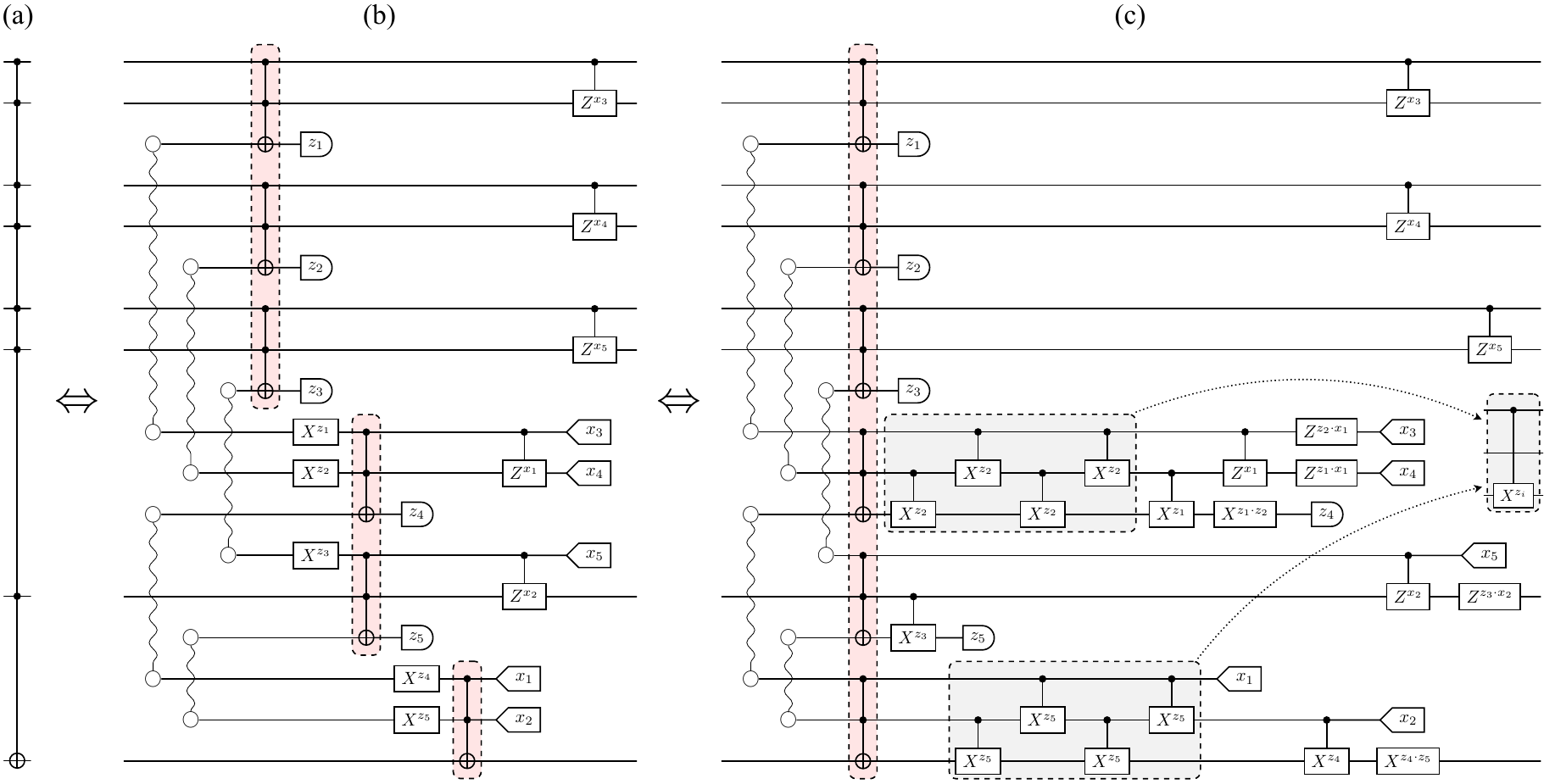}
\caption{Teleportation-based decomposition of an $\textrm{MCT}_8$ gate in the DQC paradigm. Toffoli gates applied on closest-neighboring qubits are indicated with a red dashed box. In panel (a) the $\textrm{MCT}_8$ gate is depicted as a unitary operation. Panel (b) depicts the teleportation-based decomposition of the $\textrm{MCT}_8$ gate that includes ten ancilla qubits. The Toffoli depth of this circuit is equal to three due to the measurement outcomes required for the conditional gates. Panel (c) depicts the teleportation-based decomposition of the $\textrm{MCT}_8$ gate with Toffoli depth equal to one after classically conditioned gates have been commuted to the right hand-side of the circuit. In the gray box it is shown how four consecutive gates can be merged to one in case the three qubits are considered all closest-neighboring to one another.}
\label{fig:decompositions}
\end{figure*}

For the $\textrm{MCT}_8$ gate, we obtain the following sequence:
\alsub{}{
i = 1: \quad n_1 = 7 \quad \longmapsto  \quad m_1 = 3, \quad \ell_1 = 1 \,, \nonumber \\
i = 2: \quad n_2 = 4 \quad \longmapsto  \quad m_2 = 2, \quad \ell_2 =0 \,. \nonumber 
}
The corresponding circuit for this decomposition is depicted in Fig.~\ref{fig:decompositions}.

\subsection{Toffoli Depth}

By construction, the recursive use of the teleportation protocol introduces new ancilla qubits, but the MCT gates are always applied on disjoint sets of qubits. The classically conditioned gates that appear in the decomposition explicitly depend on the measurement outcomes. Therefore, they impose a causal structure that requires a certain circuit depth for their implementation. Nevertheless, given that all such classically conditioned operations can be commuted to the end of the circuit, every Toffoli gate can be placed within a single layer, thereby reducing the total Toffoli depth to one, regardless of the number of controls of the MCT gate. 

The resulting optimized circuit for the $\textrm{MCT}_8$ gate is shown in Fig.~\ref{fig:decompositions}(c). Note that the circuit is decomposed in such a way that every gate is applied to closest-neighboring qubits. If, for a particular platform, the two-qubit gates can be applied across all qubits of the triplet, then four consecutive gates, inside the grey box, can be reduced to one. In Fig.~\ref{fig:toffoli_depth}(a) we compare the Toffoli depth of this decomposition with other state-of-the-art decompositions. 

\begin{figure*}[t!]
\centering
\includegraphics[width=\textwidth]{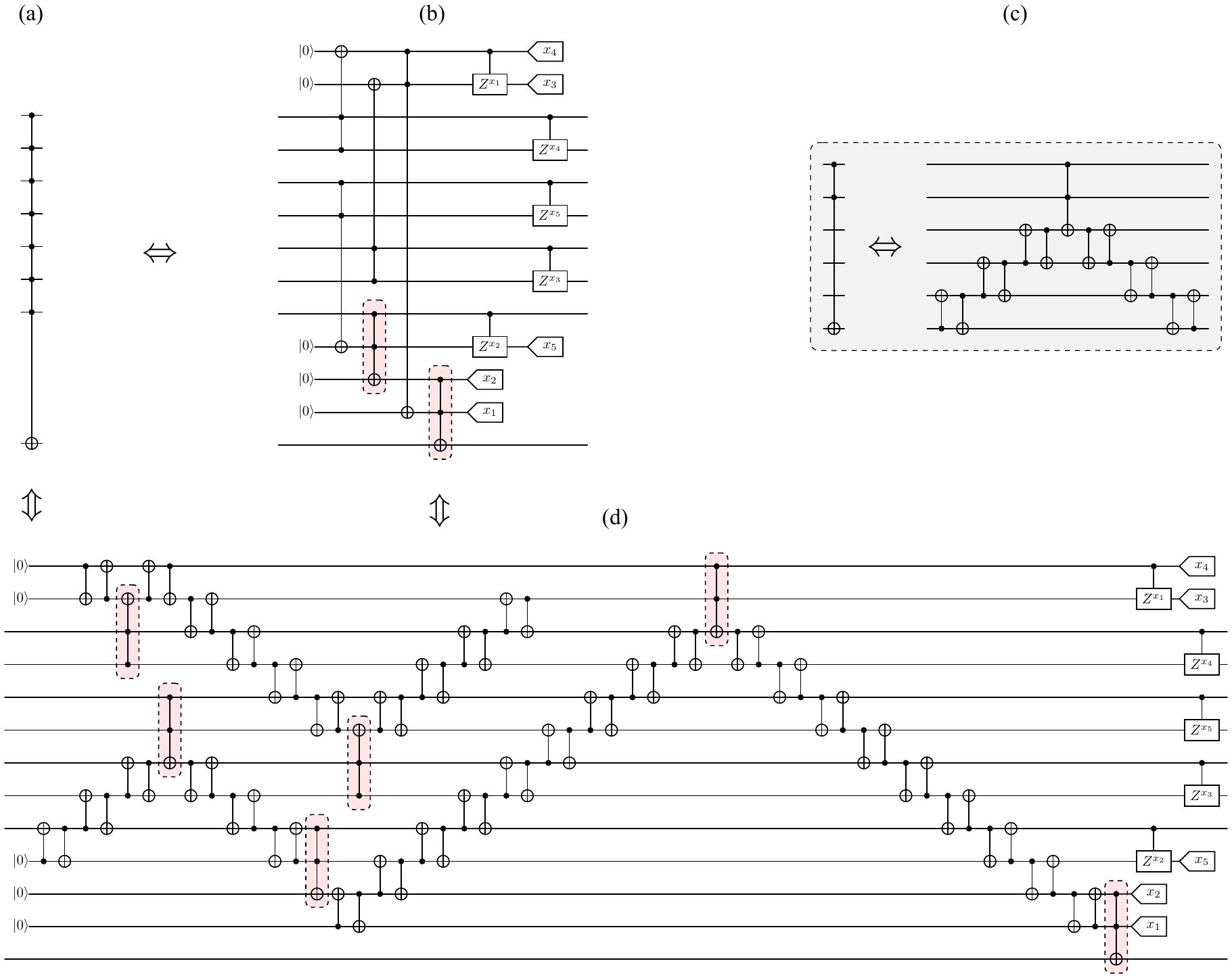}
\caption{Dutta \textit{et al.} decomposition of an $\textrm{MCT}_8$ gate in the MQC paradigm~\cite{Dutta_etal_PRA_25}. Toffoli gates applied on closest-neighboring qubits are placed within a red dashed box. Panel (a) depicts the $\textrm{MCT}_8$ gate as a unitary operation. In panel (b) the Dutta \textit{et al.} decomposition~\cite{Dutta_etal_PRA_25} of the $\textrm{MCT}_8$ gate is depicted, including the measurement-based uncomputation part, that corresponds to the lower bound (purple line) in Fig.~\ref{fig:toffoli_depth}(a). This decomposition has a Toffoli-depth equal to three, five ancilla qubits, and only two of the Toffoli gates are applied on closest-neighboring qubits. Panel (d) depicts the corresponding circuit where all Toffoli gates have been replaced by circuit primitives, shown in panel (c), that allow for only closest-neighboring interactions based on Ref.~\cite{Cruz_Murta_APLQ_24}.}
\label{fig:dutta_decomposition}
\end{figure*}

\subsection{Toffoli Count}

The Toffoli count of this decomposition is given by $1 + \sum_{i=1}^{i_{\text{max}}} m_i$, which, as proved in Appendix \ref{appB}, can be simplified to the following expression
\eq{toffoli_count}{
n - 1\,.
}
This expression is derived by observing that in each subsequent iteration we have $m_i$ Toffoli gates plus an $\textrm{MCT}_{m + \ell + 1}$ gate. The $\textrm{MCT}_{m + \ell + 1}$ gate has either more controls than a Toffoli gate, so the protocol continues to the next iteration, or it is indeed a Toffoli gate, so the protocol terminates, and thus this last Toffoli gate is taken into account by adding an extra factor of one. 

In Fig.~\ref{fig:toffoli_depth}(b) we compare the Toffoli count for different decompositions. We observe that the Toffoli count increases linearly for all decompositions, but the teleportation-based decomposition requires fewer Toffoli gates than the other decompositions for any MCT gate with more than five control qubits. 

\subsection{Entangled Pairs and Ancilla Count}
Given that $2m_i$ is the number of measurements per teleportation, the number of entangled pairs required for the protocol is equal to $\sum_{i=1}^{i_{\text{max}}}  m_i$ which as proved in Appendix \ref{appB} can be simplified to
\eq{}{
n-2 \,,
}
while the ancilla qubit count is twice that value, i.e.,
\eq{}{
2n - 4 \,.
}

In Fig.~\ref{fig:toffoli_depth}(c) we compare the ancilla count of the teleportation-based decomposition with other state-of-the-art decompositions. For the former we observe a linear increase, while for the latter a constant number of ancilla qubits is needed, which is also quite low in value.

\section{Error Analysis of Decompositions}
\label{sec_comparison}

The teleportation-based decomposition requires entanglement to be distributed before the Toffoli layer. In the MQC paradigm, which involves a single QPU, this distribution entails routing locally generated entangled qubits to the appropriate qubits using SWAP gates. Such routing is expected to contribute substantially to the overall two-qubit-gate depth of the decomposition. By contrast, in a DQC paradigm, where multiple QPUs are interconnected through entanglement links, the required entangled states can be distributed photonically among the QPUs without increasing the circuit depth. Establishing entanglement across QPUs introduces its own sources of noise and operational overhead. Nevertheless, DQC appears to provide the natural setting in which the depth advantage of the teleportation-based decomposition can be leveraged.

In this section, we investigate, under the same noise model, the performance of the $\textrm{MCT}_8$ gate implemented within the MQC and DQC paradigms. For the MQC paradigm, we employ the state-of-the-art decomposition of Dutta \textit{et al.} that achieves the lower bound of Toffoli depth derived in Ref.~\cite{Dutta_etal_PRA_25} [see purple line in Fig.~\ref{fig:toffoli_depth}(a)], which is optimal for this case. For the DQC paradigm, we employ the teleportation-based decomposition proposed in this work. The comparison therefore effectively determines which of the two computational paradigms is preferable as a function of the relative noise affecting local Toffoli gates and distributed entanglement links.

In order to provide a fair comparison, we assume that regardless of the computing paradigm within each QPU (that would be each triplet of qubits for the teleportation-based decomposition) gates can be applied only on closest-neighboring qubits which are serially connected. For the teleportation-based decomposition that is shown in Fig.~\ref{fig:decompositions}(c). For the Dutta \textit{et al.} decomposition we substitute each Toffoli gate with the circuit that implements the same operation using only gates applied on closest-neighboring qubits, based on an approach found in Ref.~\cite{Cruz_Murta_APLQ_24} and shown in Fig.~\ref{fig:dutta_decomposition}(c). The full circuit is presented in Fig.~\ref{fig:dutta_decomposition}(d). We note that a grid-type connectivity within each QPU would be beneficial to both decompositions. However, we adopt a one-dimensional nearest-neighbor connectivity in order to keep the comparison relatively simple, since the main conclusion remains unchanged.

\begin{figure*}[t!]
\centering
\includegraphics[width=\textwidth]{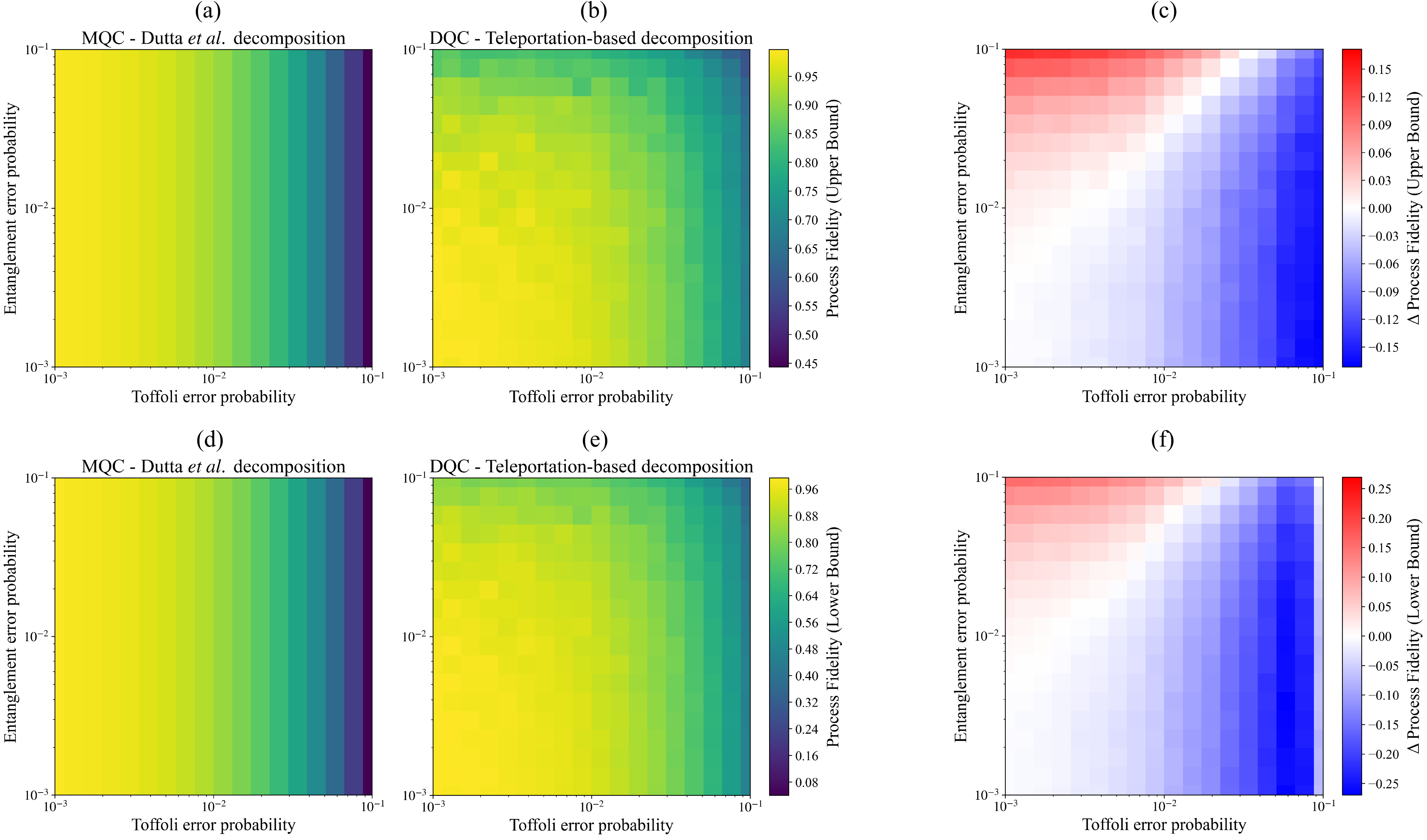}
\caption{Panels (a) and (b) depict side by side the upper bound of process fidelity of the Dutta \textit{et al.} decomposition in the MQC paradigm and the teleportation-based decomposition in the DQC paradigm, respectively. Both are plotted against the Toffoli error rate and the entanglement error rate that range from $10^{-3}$ to $10^{-1}$. Panel (c) depicts a value called $\Delta$ process fidelity, corresponding to the subtraction of the fidelity found for the teleportation-based decomposition from the Dutta \textit{et al.} decomposition. Red is the region where the Dutta \textit{et al.} decomposition in the MQC paradigm is preferable and blue the region where the teleportation-based decomposition in the DQC paradigm is preferable. Analogously panels (d) and (e) depicts to the lower bound of the process fidelity for each decomposition, and panel (f) depicts the corresponding $\Delta$ process fidelity. Numerical simulations were carried out using the \texttt{Qiskit} software framework~\cite{qiskit}.}
\label{fig:error_analysis}
\end{figure*}

We assume that quantum gates are affected by depolarizing noise. A depolarizing channel transforms a quantum state $\qs$ as follows
\eq{}{
\qs \quad \longmapsto  \quad (1-p) \qs + p \frac{\m{I}}{2^d} \,, 
}
where $\m{I}$ is the identity operator, which has the same dimensions as $\qs$, i.e., $d$. The error probability is represented by $p$. The same type of noise is assumed to affect entangled states which implies that after creation and distribution they are are Werner states with entanglement fidelity equal to $1 - 3p/4$~\cite{NielsenChuang}. Finally, qubit initializations and read-out measurements are assumed to be affected by bit-flip errors, represented by the following single-qubit channel
\eq{}{
\qs \quad \longmapsto  \quad (1-p) \qs + p \m{X} \qs \m{X}^\dag \,,
}
where $\m{X}$ is the NOT gate.

The noisy circuits correspond to quantum channels, whose process fidelity~\cite{Raginsky_PLA_01, Gilchrist_Langford_Nielsen_PRA_05}, $\f{F}_{\text{pro}}$, quantifies how close they are to the corresponding noiseless circuits. For multiple-qubit circuits, where a direct calculation of $\f{F}_{\text{pro}}$ is challenging, the process fidelity can be upper and lower bounded by the following expression~\cite{Hofmann_PRL_05}
\eq{lower_bound}{
\f{F}_z + \f{F}_c - 1 \leq \f{F}_{\text{pro}} \leq \min\{ \f{F}_z, \f{F}_c \} \,,
} 
where $\f{F}_z$ and $\f{F}_c$ are classical fidelities corresponding to two complementary bases.

Consider for example an $\textrm{MCT}_{n+1}$ gate that maps the set of eigenstates in the Z basis $\{ \ketf{\psi_i^\text{in}} \}_{i=0}^{2^{n+1} - 1}$ into another set of orthonormal states $\{ \ketf{\psi_i^\text{out}} \}_{i=0}^{2^{n+1} - 1}$, i.e.,
\eq{}{
\textrm{MCT}_{n+1} \ketf{\psi_i^\text{in}} = \ketf{\psi_i^\text{out}} \,.
}
Then, we can estimate the classical fidelity in the Z basis using the expression
\eq{}{
\f{F}_z = \frac{1}{2^{n+1}} \sum_{i=0}^{2^{n+1} - 1} \prob(\psi_i^\text{out} | \psi_i^\text{in}) \,,
}
where $\prob(\psi_i^\text{out} | \psi_i^\text{in})$ is the probability of getting the correct output state for a given input state.

Analogously, the classical fidelity in the complementary basis is given by
\eq{}{
\f{F}_c = \frac{1}{2^{n+1}} \sum_{i=0}^{2^{n+1} - 1} \prob(\varphi_i^\text{out} | \varphi_i^\text{in}) \,,
}
where $\{ \ketf{\varphi_i^\text{in}} \}_{i=0}^{2^{n+1} - 1}$ is a set of eigenstates in a complementary to the Z basis, i.e., $|\braketf{\varphi_i^\text{in}}{\psi_j^\text{in}}|^2 = 1/2^{n+1} \, \forall \, i,j \in \{ 0, \cdots , 2^{n+1} - 1\}$. An example of such a basis is the one created via the quantum Fourier transformation~\cite{Kunz_IEEE_79, Ekert_Jozsa_RMP_96} on the eigenstates of the Z basis, i.e., $\ketf{\varphi_i^\text{in}} = \textrm{QFT} \ketf{\psi_i^\text{in}}$. 

The results are presented in Fig.~\ref{fig:error_analysis}. Given the multiple error probabilities involved, we adopt a simplified error model. We vary entanglement and Toffoli error probability from $10^{-3}$ to $10^{-1}$ by dividing this range into 18 equal parts, and assume a hierarchical scaling in which two-qubit gate errors are one order of magnitude smaller than Toffoli errors, and single-qubit gate errors are two orders of magnitude smaller. Initialization errors are taken to match the single-qubit gate error rates, while readout errors are taken to match the two-qubit gate error rates. The idling error that impacts the qubits when they need to stay coherent during a computation is not taken into account as it varies significantly in the various quantum computing platforms.

In Figs.~\ref{fig:error_analysis}(a) and (b), we present the upper bound of the process fidelity for each computational paradigm. For the MQC paradigm the Dutta \textit{et al.} decomposition is employed, while for the DQC the teleportation-based decomposition. For each decomposition the process fidelity is computed from 200 shots for each of the $2^8$ input states. For the former entanglement error probability is irrelevant, but we keep it as a parameter for consistency of the presentation. In Fig.~\ref{fig:error_analysis}(c) we subtract the fidelity that corresponds to the teleportation-based decomposition from the Dutta \textit{et al.} decomposition, a value that we call $\Delta$ process fidelity. Thus, in red we have the regions where fidelity is larger for the latter and in blue the regions where fidelity is larger for the former decomposition. Analogously, in Figs.~\ref{fig:error_analysis}(d) and (e), we present the lower bound of the process fidelity for each case, and in Fig.~\ref{fig:error_analysis}(f) the corresponding $\Delta$ process fidelity.

We observe that there exist two distinct regions that indicate which implementation is preferable depending on the noise. More specifically, when Toffoli error probability is smaller than entanglement error probability, then the Dutta \textit{et al.} decomposition in the MQC paradigm is preferable, while when Toffoli error probability is larger than entanglement error probability, the teleportation-based decomposition in the DQC paradigm is preferable. In the supplemental material an alternative error model is considered for the error analysis of the two decompositions, but the results are completely analogous.

\section{Applications of MCT gates}
\label{Sec:Applications}
In this section, we outline several applications of MCT gates in quantum computing.

\subsection{Adder Operator}

The adder operator, also known as incrementor or shift operator, is defined as~\cite{Vedral_Barenco_Ekert_PRA_96, Douglas_Wang_PRA_09} 
\eq{}{
\m{A} = \sum_{i \in [0, 2^q - 1]} \ketbra{(i+1)\bmod{2^q}}{i} \,,
}
which performs an addition of 1 to the index of the ket modulo $2^q$, where $q$ is the number of register qubits. The adder operator serves as a foundational building block for constructing mathematical operators in quantum computing. For example, it can be used to perform operations such as the gradient, the divergence, and the Laplacian, and thus it plays a prominent role in variational quantum algorithms~\cite{Lubasch2020, Jaksch2023, Tomesh_etal_Q_24, Umer2025, Over_etal_CF_25}. It is also an essential subroutine in a range of other quantum algorithms, including circuit-level realizations of discrete-time quantum walks~\cite{Georgopoulos2021, Nandi2025}, lattice-gauge Hamiltonian simulations for the Schwinger model~\cite{Shaw2020}, distance-based machine-learning classification protocols~\cite{Li2022}, and non-Clifford resource–aware constructions in fault-tolerant quantum computation~\cite{Beverland2020}, among others.

\begin{figure}[b!]
\centering
\includegraphics[width=\columnwidth]{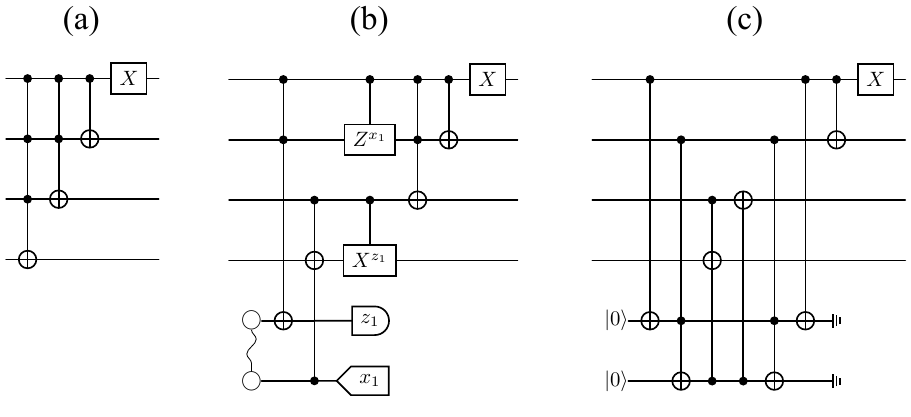}
\caption{Panel (a) depicts the MCT-based construction of the adder operator modulo $2^{4}$. Panel (b) depicts an analogous construction but substitutes the $\textrm{MCT}_4$ gate with the teleportation-based decomposition according to the protocol discussed in section~\ref{sec_tele_decomposition}. Panel (c) depicts the adder operator modulo $2^{4}$ based on Vedral \textit{et al.}~\cite{Vedral_Barenco_Ekert_PRA_96}.}
\label{fig:Adder_Circuit}
\end{figure}

The unitary implementation of the adder operator admits two formally distinct circuit constructions. In the MCT-based construction~\cite{Douglas_Wang_PRA_09}, shown in Fig. \ref{fig:Adder_Circuit}(a), a sequence of MCT gates conditionally flips each target bit only when all lower-significance bits satisfy the carry-trigger condition, thus realizing full binary carry propagation as a global higher-order control structure. An alternative construction, depicted in Fig. \ref{fig:Adder_Circuit}(c), uses only CNOT and Toffoli gates, at the expense of adding ancilla qubits~\cite{Vedral_Barenco_Ekert_PRA_96}. In this approach, sum and carry operations are handled separately: CNOT gates implement local modulo-$2$ additions, while Toffoli gates generate and propagate carry information between neighboring bit positions. 

Based on the results of previous works discussed in section~\ref{sec_decompositions} and the results of the present work discussed in section~\ref{sec_tele_decomposition}, the MCT-based construction of the adder operator can be re-expressed by decomposing each MCT gate. One possible implementation uses the decomposition proposed by Dutta \textit{et al.}, specifically the version that achieves the lower bound derived in that work. Alternatively, the teleportation-based decomposition developed in this work can be employed, as depicted in Fig.~\ref{fig:Adder_Circuit}(b).

\begin{figure}[t!]
\centering
\includegraphics[width=\columnwidth]{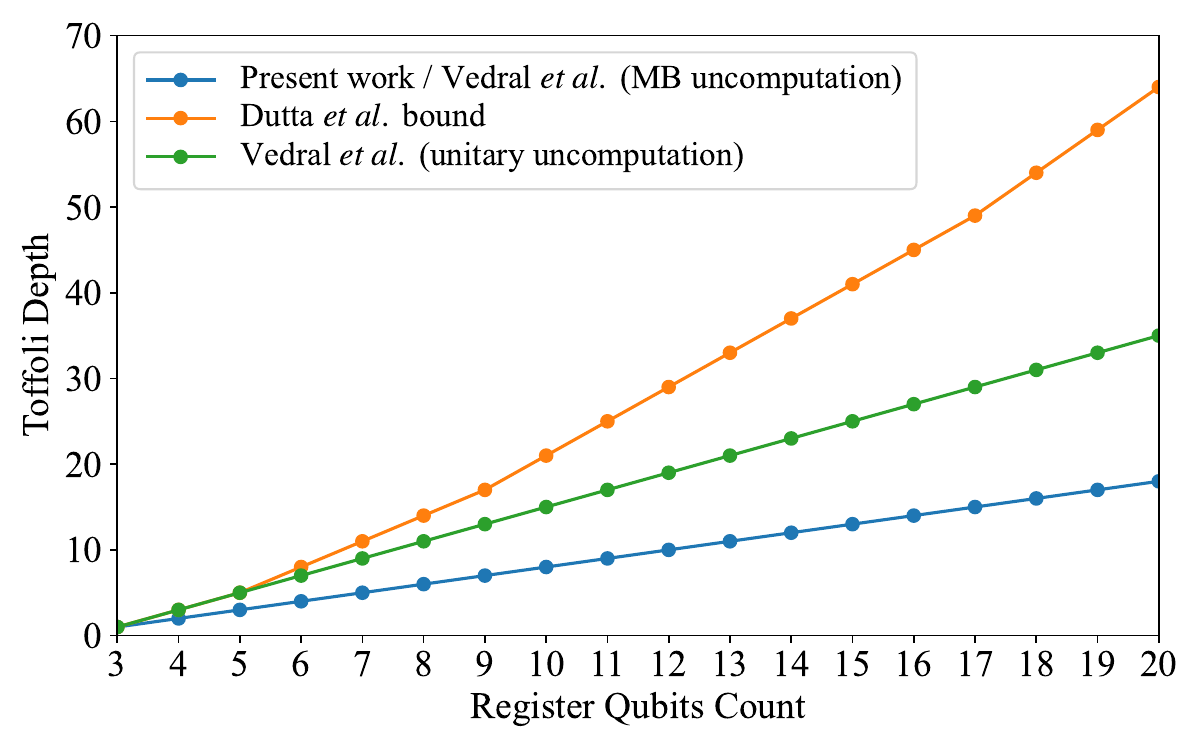}
\caption{The Toffoli depth of the adder operator is plotted for four different decompositions against the total number of register qubits. With orange the Dutta \textit{et al.} decomposition is used, with green the Vedral \textit{et al.} with unitary uncomputation, and with blue the decomposition proposed in this work. Note that the Vedral \textit{et al.} decomposition with measurement-based uncomputation achieves an equal Toffoli depth.}
\label{fig:Toffoli_depth}
\end{figure}

In Fig.~\ref{fig:Toffoli_depth} we compare the aforementioned two approaches and also consider the Vedral \textit{et al.} implementation~\cite{Vedral_Barenco_Ekert_PRA_96} of the adder operator as a whole. The teleportation-based decomposition and the Vedral \textit{et al.} decomposition with measurement-based uncomputation achieve the minimum Toffoli depth for any number of register qubits. The Vedral \textit{et al.} decomposition with unitary uncomputation performs second in this optimization, while the Dutta \textit{et al.} decomposition performs worst.

\subsection{Quantum Read-Only Memory}
\label{Sec:QROM}

Quantum read-only memory (QROM) is a data-loading primitive that implements a reversible \emph{lookup} from an address register to a classical data table, writing the addressed word into a data register. In its common form, QROM realizes a unitary of the form
\eq{QROM}{
\ket{r,a,0,d} \quad \longmapsto  \quad \ket{r,a,0,d \oplus (r \cdot w_{a}) } \,,
}
where $a$ is the address, $w_{a}$ is the classical word stored at address $a$, $d$ is  the data register, and $r$ is an optional read-enable.

A standard construction discussed in Ref.~\cite{Phalak2022} is the unary iteration implementation, where the circuit naturally decomposes into three stages per word, compute, data read, and un-compute, a structure explicitly described for the naive QROM circuit. Within this unary-iteration approach, MCT gates implement the reversible address decoding needed for the compute and un-compute stages. The compute stage uses an MCT gate whose controls include the read-enable and the address lines, flipping a dedicated “CNOT control line” (work qubit) only when the address matches the currently iterated pattern. After the data CNOT gates are applied, the same MCT gate is applied again to un-compute the work qubit back to $\ket{0}$. This un-compute is functionally essential: without resetting the temporary select qubit, subsequent word-CNOT layers can be spuriously triggered, corrupting the output. In short, MCT gates are not merely incidental to these QROM circuits; they are the mechanism that (i) constructs a clean selection predicate and (ii) removes it after use, ensuring a coherent and composable memory-read primitive. 

Consider an 8-qubit layout ($r$, $a_{0}$, $a_{1}$, $a_{2}$, $s$, $d_{0}$, $d_{1}$, $d_{2}$), as shown in Fig. \ref{fig:applications}(a), and focus on a single address, $a = 000$, storing the word $w_{000} = 101$ on ($d_{0}$, $d_{1}$, $d_{2}$). The single-word unary-iteration subroutine for this address consists of: (i) Preparation: apply X-gates on ($a_{0}$, $a_{1}$, $a_{2}$ ) so that $a=000$ is mapped to $a = 111$ on the MCT gate controls, (ii) Compute: apply an ${\rm MCT}_{5}$ gate, which flips $s$ iff $r=1$ and the address matches $a = 000$, (iii) Read: apply CNOT gates controlled by $s$ to the data bits where $w_{000}$ has $1$s, (iv) Un-compute: apply the same ${\rm MCT}_{5}$ gate again to return $s$ to $\ket{0}$ state, (v) X-undo: reapply X-gates on ($a_{0}$, $a_{1}$, $a_{2}$) to restore the address register. Fig.~\ref{fig:applications}(a) realizes this mapping, where $d_{0}$ and $d_{2}$ are flipped iff $r=1$, while $d_{1}$ remains unchanged. The ${\rm MCT}_{5}$ gates appearing in Fig.~\ref{fig:applications}(a) can be realized via teleportation-based protocol, as illustrated in Fig. \ref{fig:mct5}.

In various encoding frameworks~\cite{Babbush2018, Sunderhauf2024}, a QROM-like circuit is used to inject classical-data into quantum circuits. More broadly, any encoding scheme that must condition large numbers of controlled operations on classical tables can employ the same compute–use–uncompute template, with MCT gates providing the reversible control logic. Accordingly, entanglement-assisted non-unitary construction of MCT gates provides a way for low-depth implementation of these QROM subroutines. 

\subsection{Quantum Machine-Learning}
MCT gate operations arise naturally in a range of quantum machine-learning architectures, including quantum perceptron and neuron~\cite{Tacchino2019}, and rule-based quantum decision trees~\cite{Cuellar2024}, to name a few. In particular, whenever a circuit implements a high-order conjunction of several qubits onto a designated output or ancilla qubit, an MCT gate provides a direct reversible realization of this conditional update. Therefore, MCT gates frequently appear as enabling primitives for reversible computing. In the following, we briefly discuss two such instances in which MCT gates play fundamentally different functional roles within the machine-learning architecture.

\begin{figure}[t!]
\centering
\includegraphics[width=\columnwidth]{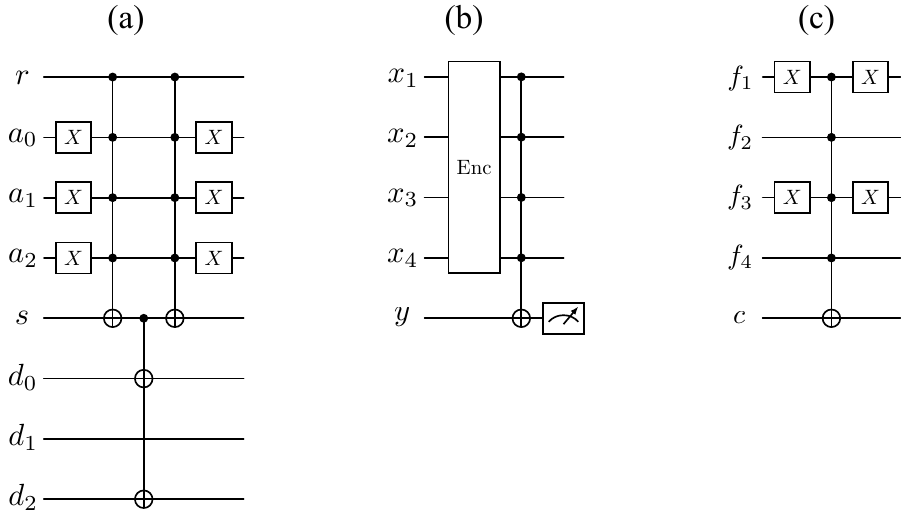}
\caption{Illustrations of MCT gates across diverse quantum circuit architectures. MCT gates are utilized in the subroutines of (a) QROM, (b) quantum neurons, and (c) quantum decision trees. Here, $r$ is a read-enable qubit, $a_{0}a_{1}a_{2}$ is the 3-bit address, $s$ is a work/selection qubit, and $d_{0}d_{1}d_{2}$ is the 3-bit data register. $x_i$ and $f_i$ are encoded data/features, and $y$ and $c$ are output or ancilla qubits.}
\label{fig:applications}
\end{figure}

The first instance is quantum perceptron/neuron constructions~\cite{Tacchino2019}, where MCT gates function as circuit-level analogues of nonlinear activation primitives. After encoding the input-weight into the amplitudes of a register, an MCT gate from the encoding qubits to a target ancilla translates the classification outcome onto that ancilla. The simplest toy example of this instance is a hard-threshold neuron that outputs $1$ only when all encoded features are $1$. For example, for encoding qubits $x_{0}$, $x_{1}$, $x_{2}$, $x_{3}$ and a single output ancilla $y$ initialized to $\ket{0}$, the circuit comprises a single ${\rm MCT}_{5}$ gate with controls on encoding qubits and target on $y$, as illustrated in Fig. \ref{fig:applications}(b). Here, the non-unitary construction of the ${\rm MCT}_{5}$ gate, as depicted in Fig. \ref{fig:mct5}, enables a shallower realization of the quantum circuit of Fig. \ref{fig:applications}(b). For a computational-basis input $\ket{x_{0}x_{1}x_{2}x_{3},y}$, the action is $\ket{x,y}~\longmapsto~\ket{x,y\oplus~x_{0}\wedge{x_{1}}\wedge{x_{2}}\wedge{x_{3}}}$, where $x_{0}\wedge{x_{1}}\wedge{x_{2}}\wedge{x_{3}} \in \{0, 1\}$. Measurement on $y$ produces the classification label. 

The second instance occurs in rule-based classifiers such as quantum decision trees~\cite{Cuellar2024}. A classical decision-tree inference step can be represented as a rule: if a conjunction of feature predicates is satisfied, then flip (or set) the class label. Here, MCT gates are used to build the inference circuit, with each MCT gate implementing an explicit decision rule for class prediction. In this setting, MCT gates arise when a rule jointly depends on multiple feature predicates, so that the predicted class bit flips if and only if all feature tests are satisfied. For example, for feature bits $f_{0}, f_{1}$, $f_{2}$, $f_{3}$ and a class qubit $c$, with $c$ initialized to $\ket{0}$, a single rule is implemented by converting the two 0-bits into 1-bits using X gates in $f_{0}$ and $f_{2}$, applying an ${\rm MCT}_{5}$ gate onto $c$, and then undoing the X gates: $U_{\rm rule} = X_{f_{0}}X_{f_{2}}{\rm MCT}_{5}X_{f_{0}}X_{f_{2}}$, as shown in Fig. \ref{fig:applications}(c). The ${\rm MCT}_{5}$ gate can be implemented via a teleportation-based non-unitary construction employing three Toffoli gates, four ancilla qubits, four mid-circuit measurements, and a small number of additional two-qubit gates, as shown in Fig. \ref{fig:mct5}.

\section{Conclusion}
\label{sec_conclusion}

In this work, we introduce a teleportation-based decomposition of the MCT gate that achieves unit Toffoli depth while requiring fewer Toffoli gates than existing state-of-the-art decompositions for more than five control qubits. The reduction in Toffoli depth comes at the cost of a number of ancilla qubits that scales linearly with the number of controls. The teleportation-based decomposition is, by construction, naturally suited to the DQC paradigm, where entanglement can be distributed across QPUs, although it can in principle also be implemented within the traditional MQC paradigm. An error analysis incorporating imperfections in both entanglement establishment and quantum gate implementation shows that, when the entanglement error probability is lower than the corresponding local Toffoli gate error probability, the teleportation-based decomposition in a DQC setting provides the preferable implementation of an MCT gate. These results highlight a promising route toward applications ranging from variational quantum algorithms to quantum memory and quantum machine learning.

\section{Acknowledgments}

This work is supported by the EU HORIZON—Project 101080085 – QCFD, the National Research Foundation, Singapore, and A*STAR under its CQT Bridging Grant. 

\appendix

\section{Proof for Eq~\eqref{eq:imax}}
\label{appA}

For an even value $n_i \geq 4$, we have $(-1)^{n_i} = 1$, which, due to Eqs.~\eqref{eq:l_expression} and \eqref{eq:m_expression}, results in $\ell_i = 0$ and $m_i = n_i/2$. For an odd value $n_i \geq 3$, we have $(-1)^{n_i} = -1$, which results in $\ell_i = 1$ and $m_i = (n_i - 1)/2$. Thus, the recursion condition in Eq.~\eqref{eq:recursion} can be re-written as
\eq{recursion_cond_2}{
n_{i + 1} = \left\lceil \frac{n_i}{2} \right\rceil \,,
}
regardless of the parity of $n_i$.

We claim that for any iteration $i \geq 1$,
\eq{iterate_closed_form}{
n_i=\left\lceil \frac{n_1}{2^{i-1}}\right\rceil = \left\lceil \frac{n}{2^{i-1}}\right\rceil .
}
The claim is trivially valid for $i=1$. Let us assume that it holds for some $i > 1$. Using the identity $\left\lceil \frac{x}{2}\right\rceil = \left\lceil \frac{1}{2}\left\lceil x\right\rceil\right\rceil$
(valid for all $x \in \set{R}$), we obtain
\eq{appendixexpression}{
n_{i+1}
=\left\lceil \frac{n}{2^i}\right\rceil
=\left\lceil \frac{1}{2}\frac{n}{2^{i-1}}\right\rceil
=\left\lceil \frac{1}{2}\left\lceil \frac{n}{2^{i-1}}\right\rceil\right\rceil
=\left\lceil \frac{n_i}{2}\right\rceil,
}
which proves \eqref{eq:iterate_closed_form} by induction.

Based on the condition in Eq.~\eqref{eq:term_condition}, the protocol terminates at $n_{i_{\text{max}}}=4$ (for an even case) or $n_{i_{\text{max}}}=3$ (for the odd case). Thus, we have 
\eq{}{
n_{i_{\text{max}}}=\left\lceil \frac{n}{2^{i_{\text{max}}-1}}\right\rceil \Rightarrow 3 \leq \left\lceil \frac{n}{2^{i_{\text{max}}-1}}\right\rceil \leq 4 \,.
}
This is equivalent to 
\eq{}{
2 < \frac{n}{2^{i_{\text{max}}-1}} \leq 4 \Leftrightarrow 2^{i_{\text{max}}} < n \leq  2^{i_{\text{max}} + 1} \,,
}
which implies
\eq{}{
i_{\text{max}} < \log_2(n) \leq i_{\text{max}} + 1 \Rightarrow i_{\text{max}} = \left\lceil \log_2(n) \right\rceil - 1 \,,
}
which completes the proof for $n \geq 3$. For $n=2$ it is straightforward to see that we only need to apply the teleportation protocol once, and thus $i_{\text{max}} = 1$.

\section{Simplification of $\sum_{i=1}^{i_{\max}} m_i$}
\label{appB}

From the parity analysis in Appendix~\ref{appB}, we have
\eq{}{
m_i = \left\lfloor \frac{n_i}{2} \right\rfloor \,,
}
Using the identity $\lfloor \frac{x}{2} \rfloor + \lceil \frac{x}{2} \rceil = x$ (valid for all $x \in \set{R}$) and Eq.\eqref{eq:appendixexpression} we get
\eq{}{
m_i = n_i - \left\lceil \frac{n_i}{2} \right\rceil = n_i - n_{i+1} \,,
}
so the sum is equal to
\eq{}{
\sum_{i=1}^{i_{\max}} m_i = n_1 - n_{i_{\max}+1} = n - n_{i_{\max}+1} = n - \left\lceil \frac{n_{i_{\max}}}{2} \right\rceil \,.
}
Given that $\left\lceil \frac{n_{i_{\max}}}{2} \right\rceil$ is equal to 2 for both the even ($n_{i_{\max}} = 4$) and odd ($n_{i_{\max}} = 3$) cases, we get
\eq{}{
\sum_{i=1}^{i_{\max}} m_i = n - 2 \,.
}

\bibliography{Bibliography/bibliography}

\pdfximage{\supplementfilename}
\def\numbersupplementpages{\the\pdflastximagepages}
\foreach \x in {1,...,\numbersupplementpages}
{\onecolumngrid\clearpage\twocolumngrid\includepdf[pages={\x}]{\supplementfilename}}

\end{document}